\documentclass{article}
\usepackage{amsmath,graphicx,hyperref}
\usepackage{amsmath,amssymb,amsfonts}
\usepackage{algorithmic}
\usepackage{graphicx}
\usepackage{textcomp}
\usepackage{multirow}
\usepackage{subcaption}
\usepackage{hyperref}
\usepackage{array}
\usepackage{adjustbox}
\usepackage{overpic}
\usepackage{xcolor}
\usepackage[utf8]{inputenc}  
\usepackage{tikz}            
\usepackage[dvipsnames]{xcolor}
\usetikzlibrary{arrows.meta, calc}
\def\ConvColor{rgb:red,1;green,1;blue,1;white,3}
\def\edgecolor{rgb:black,1}
\def\FcColor{rgb:black,1}
\def\SelfAttnColor{rgb:red,4;green,0.5;blue,0.5;white,2}
\usepackage{pgfplots}        
\usepackage{amsmath}         
\usepackage{graphicx}        
\usepackage{caption}         
\usetikzlibrary{positioning, 3d}
\usepackage{import}
\usetikzlibrary{quotes,arrows.meta}
\usetikzlibrary{positioning}

\def\edgecolor{rgb:blue,4;red,1;green,4;black,3}
\newcommand{\midarrow}{\tikz \draw[-Stealth,line width =0.8mm,draw=\edgecolor] (-0.3,0) -- ++(0.3,0);}

\usepackage{Ball}
\usepackage{Box}
\usepackage{RightBandedBox}

\def\x{{\mathbf x}}
\def\E{{\mathbb{E}}}

\newcommand{\KAKA}[1]{\textcolor{black}{#1}}

\usepackage{spconf}
\title{large-size multivariate scale-free texture generation from small-size learning sets with SELF-ATTENTION WASSERSTEIN-GAN}
\name{Jhonatan Ancco Cerda, Kinan Abbas, Stephane G. Roux and Patrice Abry\thanks{The support of the Centre Blaise Pascal IT test platform at ENS de Lyon (France) for Machine Learning facilities is gratefully acknowledged. The platform operates the SIDUS solution developed by E. Quemener \cite{sidus}.}
\address{CNRS, ENS de Lyon, LPENSL, UMR5672, 69342 Lyon Cedex 07, France\\
\{firstname.lastname\}@ens-lyon.fr}
}
\begin{document}
\maketitle
\ninept
\begin{abstract}

Multivariate scale-free texture images are nowadays central in numerous modern applications very different in nature. 
A common challenging issue consists of being able to generate such multivariate textures at any arbitrary size, from a learning set of (possibly much) smaller size.
Parametric models often turn out to be too restrictive for the generation of realistic textures, thus prompting neural network training-based generation. 
While several neural network approaches propose generation at sizes larger than training, they do not explicitly assess the quality of the generation by quantifying how well multiscale or scale-free statistics in textures are reproduced and rather rely on perceptual assessment. 
The present work combines convolutional Wasserstein-GAN architectures with self-attention layers to generate textures of arbitrary size while focusing on reproducing multiscale statistics, without coding a model into the loss function.
The relevance of using a self-attention layer is studied. 
Quantitative assessment conducted on multifractal textures, representative of a broad class of multiscale textures, over ten independent trainings shows that the proposed architecture is able to generate textures with accurate statistics for sizes $8 \times 8$ larger than the originals.
The results show that placing self-attention in the critic provides the most consistent performance across output sizes.

\end{abstract}

\begin{keywords}
Wasserstein GAN, self-attention layer, large size extrapolation, scale-free textures
\end{keywords}

\section{Introduction}
\label{sec:intro}
\noindent \textbf{Context.} Textures, i.e., images with no geometry and thus characterized by their joint statistical distributions, have always been considered essential in image processing.
Nowadays, in most recent and modern applications, with the multiplication of sensors, most textures are multivariate (several measured textures characterize the system of interest) and multiscale (their statistics follow some evolution rule with respect to the analysis scale). 
For instance, multivariate scale-free textures have statistics that are jointly covariant under dilation of the analysis scale \cite{m74}.
Classical stochastic process-based parametric models usually suffer from {not} being versatile enough for the generation of rich enough textures, often limited to Gaussian statistics, or too rigid non Gaussian models. 
This triggered massive neural network training based generation, yet not often considering situations when only small size ground truth images are available.
Further, most works do not usually explicitly seek to reproduce multiscale or scale-free non Gaussian statistics and rely mostly on perceptual rather than quantitative performance assessment. 
The present {work} aims to address such limitations. \\
\noindent \textbf{Related work.} Generative Adversarial Networks (GAN), introduced in~\cite{goodfellow2014generative}, marked a major advance in image synthesis and demonstrated strong performance in capturing geometric and semantic structures~\cite{karras2019style}.
GAN \KAKA{for texture generation} has also been explored~\cite{abry2022multivariate}.
The synthesis of spatially extended textures from observations available only on a smaller spatial support has a long history.
Classical non-parametric methods construct arbitrarily large realizations by resampling pixels or patches from a reference texture~\cite{EfrosLeung1999}.
Neural approaches replaced explicit resampling with learned representations~\cite{Gatys2015}, while convolutional GAN permit generation with arbitrary spatial dimensions~\cite{Ulyanov2016,Bergmann2017,beroud2023wassertein}.
Most of these works, however, rely on local or fixed-scale statistics~\cite{Shocher2019,Shaham2019} and are thus not well suited to reproducing multiscale statistics.

Recently, diffusion models have renewed interest in generation beyond the spatial dimensions of the training set \cite{Kulikov2023,He2024,Du2024}. 
However, much of such works focus on geometric images, do not consider multiscale textures (see a contrario \cite{abbas2025diffusion}) and rely on per component (rather than multivariate) perceptual quality, semantic consistency, and avoidance of duplicated objects.\\
\noindent\textbf{Goals, Contributions, and Outline.}
The goal of this work is to design and assess \KAKA{fully convolutional Wasserstein-GAN architectures trained on images of size $N_T\times N_T$ and able to generate multivariate scale-free textures of arbitrary size $N\times N$, with $N>N_T$}. 
\KAKA{The main contribution is to assess the ability of these architectures to preserve multiscale and cross-component statistical properties when generating textures larger than those used for training, and to study the impact of placing self-attention in the generator, in the critic, or in both networks.}
The proposed \KAKA{architectures are} detailed in Section~\ref{sec:ldm}.

\KAKA{In this work, multivariate multifractal textures are used as reference models for scale-free textures because they exhibit strongly non-Gaussian statistics, rich component-wise distributions, and complex inter-component dependencies.}
\KAKA{They are not considered here for their specific applications, but rather as well-documented reference models for multivariate multiscale textures.}
Their definitions and analysis are recalled in Section~\ref{sec:MF}.
Their statistical properties are assumed not known and not involved in the loss used for training. 
Section~\ref{sec:Eval} \KAKA{presents the assessment protocol and compares the statistical performance of the different self-attention placements.}

\section{Self-Attentive WGAN}
\label{sec:ldm}
\noindent\textbf{Wasserstein GAN.}
GAN train a generator $G$ to transform a random input $z$ into synthetic samples, while a discriminator learns to distinguish generated samples from training samples ~\cite{goodfellow2014generative}. 
Wasserstein GAN (WGAN) replace the discriminator with a critic $C$, known to provide a more stable training framework ~\cite{arjovsky2017wasserstein,gulrajani2017improved}. 
For training the critic, a loss function entailing a gradient penalty enforcing the Lipschitz constraint is used, 
\begin{equation}
\begin{aligned}
\mathcal{L}_C
&= -\mathbb{E}_{x\sim p_{\mathrm{data}}}[C(x)]
   + \mathbb{E}_{z\sim p_z}[C(G(z))] \\
&\quad + \KAKA{\lambda_{\mathrm{GP}}}\,\,
   \mathbb{E}_{\tilde y\sim p_{\tilde y}}
   \left[
   \max\!\left(0,\|\nabla_{\tilde y}C(\tilde y)\|_2-1\right)^2
   \right],
\end{aligned}
\end{equation}
\KAKA{where $\tilde y$ denotes an interpolation between real and generated samples, and $\lambda_{\mathrm{GP}}$ controls the gradient penalty.} 
\KAKA{The generator is trained to maximize the critic score assigned to generated samples, or equivalently to minimize}
\begin{equation}
\mathcal{L}_G
=
-\mathbb{E}_{z\sim p_z}\left[C(G(z))\right].
\end{equation}

\noindent\textbf{Spatial-support extrapolation.}
The classical WGAN architecture suffers from a major restriction:
the fully connected layer, used as the first layer of the generator, imposes that generated textures have one same and unique size, that of the training set.
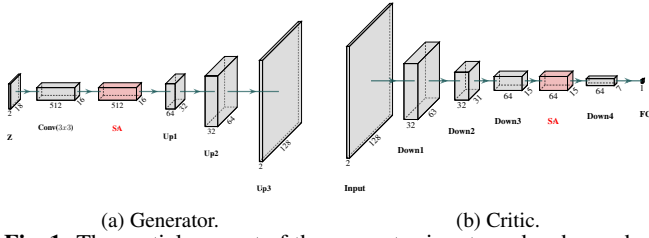
\begin{figure}[t]
\centering
\begin{subfigure}{0.48\columnwidth}
    \centering
    \resizebox{\linewidth}{!}{\resizebox{0.2\paperwidth}{!}{%
\begin{tikzpicture}

\tikzstyle{connection}=[ultra thick, every node/.style={sloped,allow upside down}, draw=\edgecolor, opacity=0.7, line width=0.8mm]
\tikzstyle{copyconnection}=[ultra thick, every node/.style={sloped,allow upside down}, draw={rgb:blue,4;red,1;green,1;black,3}, opacity=0.7, line width=0.8mm]

\pic[shift={(0,0,0)}, line width=0.4mm] at (0,0,0) 
    {Box={
        name=noise0,
        caption=\textbf{Z},
        xlabel={{2, }},
        zlabel=18,
        fill=\ConvColor,
        height=4*2,
        width=0.3*2,
        depth=4*2
    }};

\pic[shift={(1.5,0,0)}, line width=0.4mm] at (noise0-east) 
    {Box={
        name=conv0,
        caption=\textbf{Conv($3x3$)},
        xlabel={{512, }},
        zlabel=16,
        fill=\ConvColor,
        height=2*2,
        width=6*2,
        depth=2*2
    }};
\draw [connection] (noise0-east) -- node {\midarrow} node[above, midway, sloped=false, font=\bfseries] {} (conv0-west);

\pic[shift={(1.5,0,0)}, line width=0.4mm] at (conv0-east) 
    {Box={
        name=attn,
        caption=\textcolor{red}{SA},
        xlabel={512,},
        zlabel=16,
        fill=\SelfAttnColor,
        height=2*2,
        width=6*2,
        depth=2*2
    }};

\draw [connection] (conv0-east) -- node {\midarrow} node[above, midway, sloped=false, font=\bfseries] {} (attn-west);

\pic[shift={(2.0,0,0)}, line width=0.4mm] at (attn-east) 
    {Box={
        name=conv1,
        caption=\textbf{Up1},
        xlabel={{64, }},
        zlabel=32,
        fill=\ConvColor,
        height=4*2,
        width=3,
        depth=4*2
    }};
\draw [connection] (attn-east) -- node {\midarrow} node[above, midway, sloped=false, font=\large\bfseries] {} (conv1-west);

\pic[shift={(2.2,0,0)}, line width=0.4mm] at (conv1-east) 
    {Box={
        name=conv2,
        caption=\textbf{Up2},
        xlabel={{32, }},
        zlabel=64,
        fill=\ConvColor,
        height=8*2,
        width=2*2,
        depth=8*2
    }};
\draw [connection] (conv1-east) -- node {\midarrow} node[above, midway, sloped=false, font=\large\bfseries] {} (conv2-west);

\pic[shift={(3.25,0,0)}, line width=0.4mm] at (conv2-east) 
    {Box={
        name=conv3,
        caption=\textbf{Up3},
        xlabel={{2, }},
        zlabel=128,
        fill=\ConvColor,
        height=16*2,
        width=0.5*2,
        depth=16*2
    }};
\draw [connection] (conv2-east) -- node {\midarrow} node[above, midway, sloped=false, font=\large\bfseries] {} (conv3-west);

\end{tikzpicture}
}%
}
    \caption{Generator.}
\end{subfigure}
\hfill
\begin{subfigure}{0.48\columnwidth}
    \centering
    \resizebox{\linewidth}{!}{\resizebox{0.20\paperwidth}{!}{%
\begin{tikzpicture}
\tikzstyle{connection}=[ultra thick, every node/.style={sloped,allow upside down}, draw=\edgecolor, opacity=0.7, line width=0.8mm]

\pic[shift={(0,0,0)}, line width=0.4mm] at (0,0,0) 
    {Box={
        name=input0,
        caption= \textbf{Input},
        xlabel={{2, }},
        zlabel=128,
        fill=\ConvColor,
        height=16*2,       
        width=0.4*2,
        depth=16*2
        }
    };

\pic[shift={(2.5,0,0)}, line width=0.4mm] at (input0-east) 
    {Box={
        name=down1,
        caption= \textbf{Down1},
        xlabel={{32, }},
        zlabel=63,
        fill=\ConvColor,
        height=8*2,
        width=2*2,
        depth=8*2
        }
    };
\draw [connection] (input0-east) -- node {\midarrow} (down1-west);

\pic[shift={(1.8,0,0)}, line width=0.4mm] at (down1-east) 
    {Box={
        name=down2,
        caption= \textbf{Down2},
        xlabel={{32, }},
        zlabel=31,
        fill=\ConvColor,
        height=4*2,
        width=2*2,
        depth=4*2
        }
    };
\draw [connection] (down1-east) -- node {\midarrow} (down2-west);

\pic[shift={(1.3,0,0)}, line width=0.4mm] at (down2-east) 
    {Box={
        name=down3,
        caption= \textbf{Down3},
        xlabel={{64, }},
        zlabel=15,
        fill=\ConvColor,
        height=2*2,
        width=4*2,
        depth=2*2
        }
    };
\draw [connection] (down2-east) -- node {\midarrow} (down3-west);

\pic[shift={(1.0,0,0)}, line width=0.4mm] at (down3-east) 
    {Box={
        name=attn,
        caption= \textcolor{red}{SA},
        xlabel={{64, }},
        zlabel=15,
        fill=\SelfAttnColor,
        height=2*2,
        width=4*2,
        depth=2*2
        }
    };
\draw [connection] (down3-east) -- node {\midarrow} (attn-west);

\pic[shift={(1.0,0,0)}, line width=0.4mm] at (attn-east) 
    {Box={
        name=down4,
        caption= \textbf{Down4},
        xlabel={{64, }},
        zlabel=7,
        fill=\ConvColor,
        height=1*2,
        width=4*2,
        depth=1*2
        }
    };
\draw [connection] (attn-east) -- node {\midarrow} (down4-west);

\pic[shift={(1.5,0)}, line width=0.4mm] at (down4-east) 
    {Box={
        name=fc,
        caption= \textbf{FC},
        xlabel={{1, }},
        zlabel= ,         
        fill=\FcColor,
        height=1,
        width=0.5,
        depth=1
        }
    };
\draw [connection] (down4-east) -- node {\midarrow} (fc-west);

\end{tikzpicture}%
}}
    \caption{Critic.}
\end{subfigure}
\vskip-5mm
\caption{\KAKA{
The spatial support of the generator input can be changed at inference.
Red blocks indicate the two candidate self-attention (SA) locations evaluated in this work: after the first generator convolution and between the third and fourth critic convolutions.}}
\label{fig:wgan_architecture}
\end{figure}











For univariate unidimensional time series, it was proposed in \cite{beroud2023wassertein} to replace that fully-connected input layer by a convolutional layer.
This results in a
\KAKA{convolutional} generator for which the spatial support of the input noise can be changed at inference time:
The learned parameters can be used to generate samples on supports larger than those available during training. 

We extend this principle here from one-dimensional univariate time series to two-dimensional multivariate textures.
The present work explores the potential critical impact of the convolutional kernel size in the performance of the proposed generator. 
Compared performance are reported and discussed in Section~\ref{sec:Eval}.\\
\noindent\textbf{Self-attention layer.}
In addition, we devise richer generator/critic architectures by incorporating self-attention layers. 
While convolutional layers combine information within local neighborhoods, self-attention layers combine \KAKA{learned features from different spatial locations, enabling nonlocal interactions}.
Given a feature map $\mathbf F$, three learned $1\times1$ convolutions produce query, key, and value matrices $\mathbf Q$, $\mathbf K$, and $\mathbf V$. The self-attention output is
\begin{equation}
\widetilde{\mathbf F}
=
\mathbf F+
\KAKA{\alpha}\,
\operatorname{softmax}
\left(
\frac{\mathbf Q\mathbf K^\top}{\sqrt d}
\right)\mathbf V,
\end{equation}
where $d$ is the query/key dimension and \KAKA{$\alpha$} is a learned scalar.

In the generator, self-attention is inserted immediately after the first convolutional layer, \KAKA{where the spatial support is smallest during training.}
In the critic, it is inserted between the third and fourth convolutional layers, where the features combine information from broader receptive fields.
\KAKA{These positions limit the quadratic cost of attention while allowing nonlocal information to be processed before the subsequent generator or critic layers}~\cite{vaswani2017attention,zhang2019self}.\\
\noindent\textbf{Proposed architecture.}
The common backbone combines a convolutional generator and a convolutional critic, as illustrated in Fig.~\ref{fig:wgan_architecture}.
\KAKA{To balance representation capacity and computational cost, the channel depth progressively decreases in the generator and increases in the critic. The generator uses convolutional upsampling blocks with batch normalization and ReLU activations, whereas the critic uses stride-$(2,2)$ convolutions with leaky-ReLU activations for spatial downsampling.}
The generator and critic backbones contain $\simeq323$k and $\simeq68$k parameters, respectively, { excluding the optional self-attention blocks.} 
\KAKA{For comparison, a generator with a fully connected first layer would require approximately $3576$k parameters to generate images of size $64\times64\times 3$ ~\cite{radford2015unsupervised}.}

\KAKA{For clarity, Fig.~\ref{fig:wgan_architecture} displays the self-attention blocks in both the generator and the critic. Their separate and joint placements are assessed in Section~\ref{sec:Eval}.}
\section{Multivariate multiscale textures}
\label{sec:MF}

\noindent \textbf{Multivariate scale-free texture models.} M-variate multifractal random walks (MMRW), used as rich multiscale multivariate stochastic models for textures,  are defined as \cite{m74,bdm01}: 
\[\mathbf{X}(\mathbf{x}) = \bigl(X^{(1)}(\mathbf{x}), \ldots, X^{(M)}(\mathbf{x})\bigr) \makebox{ where }\]
\begin{equation}
X^{(m)}(\mathbf{x}) = e^{\omega^{(m)}_{\lambda_m}(\mathbf{x})} \, G^{(m)}_{H_m}(\mathbf{x}) ,
\label{eq:mf}
\end{equation}
and with $\{G_{H_m}^{(m)}(\mathbf{x})\}_{m=1}^M$ and $\{\omega^{(m)}_{\lambda_m}(\mathbf{x})\}_{m=1}^M$ are jointly Gaussian zero-mean fields. 

Technically, ${G_{H_m}^{(m)}(\mathbf{x})}{m=1}^M$ denote multivariate fractional Gaussian noise (fGn), characterized by the Hurst parameters ${H_m}{m=1}^M$, which control the cross-spatial correlations~\cite{abry2022multivariate,lucas2024multivariate}:
\begin{equation}
\mathbb{E}\!\left\{ G_H^{(m)}(\mathbf{x})\, G_H^{(m')}(\mathbf{x}') \right\}
= \rho_{m,m'} \, C_H(\mathbf{x}-\mathbf{x}'),
\end{equation}
where $C_H(\cdot)$ denotes the covariance function of the univariate fGn with Hurst exponent $H = (H_m+H_m')/2$, and $\rho_{m,m'} \in [-1,1]$ defines the cross-correlation coefficients.

Expanding on univariate definition \cite{bdm01}, the $\{\omega^{(m)}_{\lambda_m}(\mathbf{r})\}_{m=1}^M$  are further defined via their auto- and cross-correlation as:
\begin{equation}
\mathbb{E}\!\left\{ \omega^{(m)}_{\lambda_m}(\mathbf{x})\, \omega^{(m')}_{\lambda_{m'}}(\mathbf{x}') \right\}
= \rho^{mf}_{m,m'} \, \lambda_m \lambda_{m'}
\log\!\left(\frac{L}{\|\mathbf{x}-\mathbf{x}'\|+1}\right),
\end{equation}
for $\|\mathbf{x}-\mathbf{x}'\| \leq L$, and zero otherwise, $\rho^{mf}_{m,m'} \in [-1,1]$ defines  cross-correlation coefficients \cite{LeonarduzziEUSIPCO2018,abry2022multivariate}.

Intuitively, the $\{G_{H_m}^{(m)}(\mathbf{x})\}_{m=1}^M$ constitute jointly Gaussian textures with algebraically decaying (cross-) correlations thus inducing long-range correlations, while the $\{e^{\omega^{(m)}}_\lambda(\mathbf{x})\}_{m=1}^M$ 
act as covariance spatially structured modulations,
inducing strong departures from joint Gaussianity in MMRW, explicitly controlled jointly by  $\lambda_{m} $ and $\rho^{mf}_{m,m'}$ while preserving scale-free spatial structures at all statistical orders. 
This is why $\rho^{mf}_{m,m'}$ is referred to as the multifractal correlation, with a slight abuse of language as it quantifies higher order statistics \cite{LeonarduzziEUSIPCO2018,abry2022multivariate}.\\
\noindent \textbf{Multiscale (wavelet-based) analysis.}
It is well-documented that scale-free and multifractal textures are well-characterized using wavelet-based multiscale representations \cite{Wendt:2007:E}. Let $\{d_{X^{(m)}}(j,\mathbf{k})\}_{m=1}^{M}$ denote the M-variate coefficients of a discrete wavelet transform computed from multivariate texture $X$, defined as inner products between each $X^{(m)}$ and dilated and translated wavelet templates at scale $2^j$ and spatial location $2^j\mathbf{k}$~\cite{Mallat1998}. 
To robustly capture scale-free behaviors at all statistical orders, hence, multifractal properties, wavelet leaders are defined as local suprema of wavelet coefficients taken over finer scales and within a spatial neighborhood: $L_{X^{(m)}}(j,\mathbf{k}) = \makebox{sup}_{j'\leq j, k' \in[k2^j, (k+1)2^j[} d_{X^{(m)}}(j',\mathbf{k'})$  \cite{Wendt:2007:E,abry2022multivariate}. 

For MMRW, the first order cumulants of the logarithm of the leaders
characterizes the univariate scale-free statistics of each component via linear behaviors as function of scales:  
\begin{equation}
\label{eq:c1}
C^{(1)}_{m}(j) = \E \log L_{X^{(m)}}(j,\mathbf{k}) = c_m^{(1,0)} + c^{1}_m \log 2^j, 
\end{equation}
$\forall m=1,\ldots,M$, with $c^{1}_m = H_m + \lambda_m^2/2$, thus mostly dominated by the correlation structure and primarily reflecting the (univariate) second-order statistics \cite{bdm01,abry2022multivariate}. 
The higher order cumulants of the logarithm of the leaders reflects the cross-higher-order statistics of the process. 
Notably, the second order cumulants characterizes the scale-free cross-statistics via a linear behaviors as function of scales: 
\begin{align}
\label{eq:c2}
C^{(2)}_{m,m'}(j) &= \makebox{ Var } \log L_{X^{(m)}}(j,\mathbf{k}) \log L_{X^{(m')}}(j,\mathbf{k}) \\
&= c_{m,m'}^{(2,0)}  + c^{2}_{m,m'} \log 2^j, 
\end{align}
$\forall m, m'=1,\ldots,M$, with $c^{2}_{m,m'} = - \rho^{mf}_{m,m'} \lambda_m \lambda_{m'}$
 quantifying cross multifractality and intermittency \cite{bdm01,abry2022multivariate}.
 
 To further quantify the cross-statistics, one usually defines the wavelet covariance function as \cite{whitcher2000wavelet,LeonarduzziEUSIPCO2018,abry2022multivariate}, with $S_{m,m'}(j) = \sum_k d_{X_m}(j,k)d_{X_{m'}}(j,k)$:
 \begin{equation}
\label{eq:rhowav}
\rho^{wav}_{m,m'}(j) = \frac{S_{m,m'}(j)}{\sqrt{S_{m,m}(j)S_{m',m'}(j)}}. 
\end{equation}
For MMRW, $\rho^{wav}_{m,m'}(j)$ should not depend on $j$ and be equal to $\rho_{m,m'}$: $ \rho^{wav}_{m,m'}(j) = \rho_{m,m'}$, $\forall j>0$, a property that can be used to assess the quality of the generated textures. 

In addition, one can compute a multifractal counterpart to the wavelet covariance function as: 
\begin{equation}
\label{eq:rhowav}
\rho^{mf}_{m,m'}(j) = \frac{C^{(2)}_{m,m'}(j)}{\sqrt{C^{(2)}_{m,m}(j)C^{(2)}_{m',m'}(j)}}. 
\end{equation}
For MMRW, $\rho^{mf}_{m,m'}(j)$ should converge to $\rho^{mf}$ in the limit of fine scales $2^j \rightarrow 0$.
While this asymptotic limit is not reachable in practice, the function $\rho^{mf}_{m,m'}(j)$ can still be used for comparisons between textures from the generated and training datasets, as a means to assess that cross dependencies beyond second-order statistics (cross-correlations) are satisfactorily reproduced and for estimating $\rho^{mf}_{m,m'}$ by averaging $\rho^{mf}_{m,m'}(j)$ across fine scales.

\section{Performance assessment}
\label{sec:Eval}
\subsection{Experimental setup}
\noindent\textbf{Training dataset.}
The dataset consists of $10,000$ independent bivariate ($M=2$) MMRW images of size $128\times128$, of which $8,000$ are used for training and $2,000$ for validation. The images are generated using the circular matrix embedding toolbox described in~\cite{helgason2011fast}.
The two components have different marginal multifractal properties, with self-similarity parameters $H=[0.8,0.6]$ and multifractality parameters $\lambda=[\sqrt{0.02},\sqrt{0.04}]$. Inter-component dependencies are set to $\rho_{1,2}=-0.4$ for second-order correlations and $\rho^{mf}_{1,2}=0.6$ for cross-multifractal dependencies.
This setting is chosen because generating MMRW with different multifractal structures for each component is more demanding than the settings considered in~\cite{abry2022multivariate,beroud2023wassertein}, since the neural network must jointly reproduce distinct component-wise multiscale properties and inter-component dependencies.
Also, the dependence structure is challenging as the second order correlation is negative (global anti-correlation) while the multifractal correlation is positive (positively correlated transient or local structures). 

\noindent\textbf{Training settings.}
\KAKA{Four configurations are considered: a WGAN without self-attention (W), self-attention in the generator only (SA-G), self-attention in the critic only (SA-C), and self-attention in both networks (SA-GC).}
All configurations use the same training settings.
We use Adam with learning rate $10^{-4}$, $\beta_{\smash{1}}=0$, and $\beta_{\smash{2}}=0.9$, a gradient penalty coefficient $\KAKA{\lambda_{\mathrm{GP}}}=10$, and a batch size of $8$.
Five critic updates are performed for each generator update.
\KAKA{All configurations} are trained for $25$ epochs.

\noindent\textbf{Evaluation protocol.}
The networks are trained using $128\times128$ textures only.
At inference, the spatial size $N_z\times N_z$ of the input noise is varied to generate textures of size $N\times N$, with $N=128,256,512,1024,$ and $2048$.
The corresponding input noise sizes are $2\times N_z\times N_z$, with $N_z=18,\KAKA{34},66,130,$ and $258$, respectively.
For each \KAKA{configuration, training run,} and output size, $25$ textures are generated.
The multiscale statistics introduced in Section~\ref{sec:MF} are computed and used to estimate $c_1$, $c_2$, $\rho$, and $\rho^{mf}$.
To account for training variability, the complete experiment is repeated over $10$ independent trainings from scratch, using the same training set and training conditions.
Bias, RMSE, and standard deviation are computed for each training and then averaged over the $10$ trainings.



\subsection{Results and Discussion}

\begin{figure}[h]
    \centering
    \includegraphics[width=0.6\linewidth]{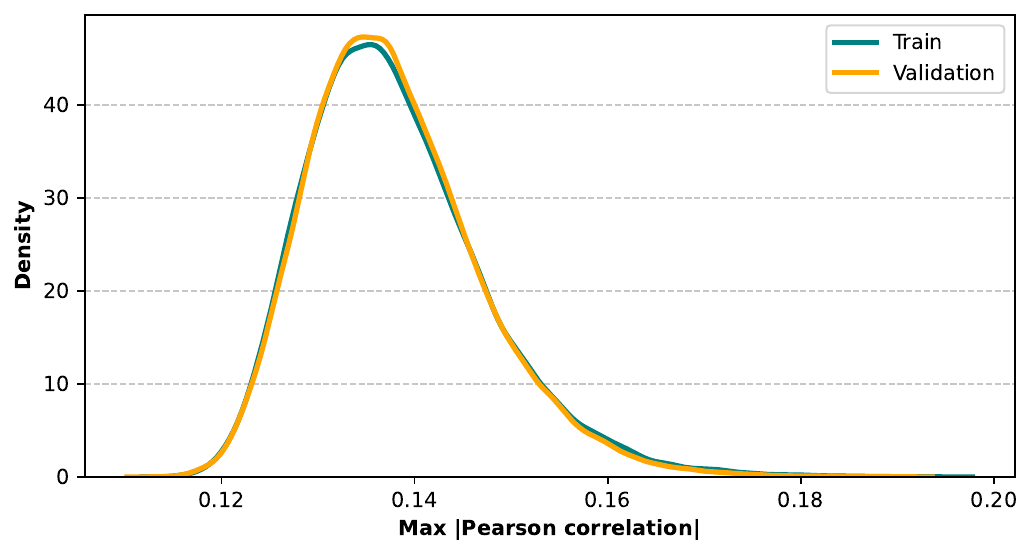}
    \vskip-5mm
    
    \caption{Distributions of the maximum absolute Pearson correlations with translated and rotated training textures, computed for \KAKA{SA-C-generated textures} and validation textures. }
    \label{fig:overfitting}
\end{figure}

\noindent\textbf{Overfitting assessment.}
To check that the generator does not simply reproduce training samples, each generated texture is compared with translated and rotated versions of the training textures, and the largest absolute Pearson correlation is retained. The same procedure is applied to the validation set. Fig.~\ref{fig:overfitting} shows nearly identical distributions, indicating no memorization.

\begin{table*}[!t]
\centering
\caption{Generation performance for the four model families (W, SA-C, SA-G, SA-CG), in terms of RMSE, Bias and standard deviations averaged across 10 independent trainings.
In addition, $V(B)$ and $V(R)$ report the variability across independent trials in terms of std of the Biases and RMSE.
\textcolor{red}{Red} indicates the overall best among all four models; \textcolor{green}{green} indicates the best among SA-C, SA-G, SA-CG when W is better; gray indicates the other two SA variants.}
\label{tab:performance}
\setlength{\tabcolsep}{1.35pt}
\renewcommand{\arraystretch}{1.02}
\resizebox{1.03\textwidth}{!}{
\begin{tabular}{cc|cccc|cccc|cccc|cccc|cccc|cccc}
\hline
& &
\multicolumn{4}{c|}{$c_1^{(1)}\,[10^{-2}]$} &
\multicolumn{4}{c|}{$c_1^{(2)}\,[10^{-2}]$} &
\multicolumn{4}{c|}{$c_2^{(1)}\,[10^{-2}]$} &
\multicolumn{4}{c|}{$c_2^{(2)}\,[10^{-2}]$} &
\multicolumn{4}{c|}{$\rho^{wav}\,[10^{-2}]$} &
\multicolumn{4}{c}{$\rho^{mf}\,[10^{-2}]$} \\
$N$ & Metric &
W & SA-C & SA-G & SA-CG &
W & SA-C & SA-G & SA-CG &
W & SA-C & SA-G & SA-CG &
W & SA-C & SA-G & SA-CG &
W & SA-C & SA-G & SA-CG &
W & SA-C & SA-G & SA-CG \\
\hline
128
& RMSE
& 10.40 & \textcolor{red}{8.43} & \textcolor{gray}{9.65} & \textcolor{gray}{9.06}
& 8.66 & \textcolor{red}{8.51} & \textcolor{gray}{11.33} & \textcolor{gray}{9.80}
& 3.54 & \textcolor{red}{2.94} & \textcolor{gray}{3.66} & \textcolor{red}{2.94}
& 4.07 & \textcolor{red}{3.52} & \textcolor{gray}{4.72} & \textcolor{gray}{4.92}
& \textcolor{red}{3.83} & \textcolor{green}{4.42} & \textcolor{gray}{5.08} & \textcolor{gray}{5.77}
& 30.8 & \textcolor{gray}{28.5} & \textcolor{gray}{32.3} & \textcolor{red}{28.2} \\
& Bias
& \textcolor{red}{-0.18} & \textcolor{gray}{-1.76} & \textcolor{gray}{1.74} & \textcolor{green}{-1.42}
& 2.74 & \textcolor{red}{2.46} & \textcolor{gray}{5.69} & \textcolor{gray}{4.38}
& 0.47 & \textcolor{gray}{-0.38} & \textcolor{gray}{0.76} & \textcolor{red}{-0.21}
& 0.74 & \textcolor{red}{0.68} & \textcolor{gray}{0.73} & \textcolor{gray}{2.74}
& \textcolor{red}{1.26} & \textcolor{green}{2.15} & \textcolor{gray}{2.46} & \textcolor{gray}{3.68}
& -28.9 & \textcolor{gray}{-26.8} & \textcolor{gray}{-30.4} & \textcolor{red}{-26.6} \\
& Std
& 8.67 & \textcolor{red}{8.13} & \textcolor{gray}{9.02} & \textcolor{gray}{8.26}
& \textcolor{red}{7.58} & \textcolor{green}{8.05} & \textcolor{gray}{8.89} & \textcolor{gray}{8.50}
& 3.52 & \textcolor{red}{2.83} & \textcolor{gray}{3.46} & \textcolor{gray}{2.94}
& 3.98 & \textcolor{red}{3.40} & \textcolor{gray}{4.28} & \textcolor{gray}{4.09}
& \textcolor{red}{3.11} & \textcolor{green}{3.64} & \textcolor{gray}{3.93} & \textcolor{gray}{4.50}
& 10.8 & \textcolor{red}{9.5} & \textcolor{gray}{10.9} & \textcolor{gray}{9.6} \\
& $V(B)$
& 6.80 & \textcolor{red}{2.17} & \textcolor{gray}{3.60} & \textcolor{gray}{4.26}
& 4.07 & \textcolor{red}{2.39} & \textcolor{gray}{6.44} & \textcolor{gray}{3.29}
& 0.76 & \textcolor{gray}{1.02} & \textcolor{gray}{1.19} & \textcolor{red}{0.58}
& 1.19 & \textcolor{red}{1.02} & \textcolor{gray}{2.60} & \textcolor{gray}{1.15}
& 2.21 & \textcolor{gray}{1.75} & \textcolor{gray}{2.33} & \textcolor{red}{1.13}
& 4.2 & \textcolor{red}{2.9} & \textcolor{gray}{4.5} & \textcolor{gray}{3.4} \\
& $V(R)$
& 2.89 & \textcolor{red}{1.05} & \textcolor{gray}{1.57} & \textcolor{gray}{2.43}
& 2.22 & \textcolor{red}{1.67} & \textcolor{gray}{4.90} & \textcolor{gray}{2.01}
& 0.77 & \textcolor{gray}{0.59} & \textcolor{red}{0.51} & \textcolor{gray}{0.58}
& 1.18 & \textcolor{red}{0.59} & \textcolor{gray}{2.31} & \textcolor{gray}{1.37}
& 1.04 & \textcolor{gray}{1.10} & \textcolor{red}{0.74} & \textcolor{gray}{1.35}
& 3.8 & \textcolor{red}{2.8} & \textcolor{gray}{4.2} & \textcolor{gray}{3.5} \\
\hline
256
& RMSE
& 5.82 & \textcolor{red}{4.17} & \textcolor{gray}{7.11} & \textcolor{gray}{5.16}
& 7.59 & \textcolor{gray}{6.92} & \textcolor{gray}{7.45} & \textcolor{red}{6.88}
& 2.19 & \textcolor{gray}{1.36} & \textcolor{gray}{1.82} & \textcolor{red}{1.30}
& \textcolor{red}{2.01} & \textcolor{green}{2.16} & \textcolor{gray}{2.25} & \textcolor{gray}{2.87}
& \textcolor{red}{3.33} & \textcolor{green}{3.40} & \textcolor{gray}{5.13} & \textcolor{gray}{6.06}
& 32.1 & \textcolor{red}{29.9} & \textcolor{gray}{34.1} & \textcolor{gray}{31.6} \\
& Bias
& 1.83 & \textcolor{red}{-0.67} & \textcolor{gray}{1.67} & \textcolor{gray}{-1.61}
& 6.03 & \textcolor{gray}{5.26} & \textcolor{gray}{4.76} & \textcolor{red}{4.62}
& 0.98 & \textcolor{gray}{0.20} & \textcolor{gray}{0.22} & \textcolor{red}{0.19}
& 0.89 & \textcolor{gray}{1.16} & \textcolor{red}{0.57} & \textcolor{gray}{2.08}
& \textcolor{red}{0.97} & \textcolor{green}{1.82} & \textcolor{gray}{2.99} & \textcolor{gray}{4.84}
& -31.4 & \textcolor{red}{-29.5} & \textcolor{gray}{-33.2} & \textcolor{gray}{-30.9} \\
& Std
& 4.49 & \textcolor{red}{3.83} & \textcolor{gray}{5.62} & \textcolor{gray}{4.33}
& \textcolor{red}{3.97} & \textcolor{green}{4.28} & \textcolor{gray}{5.20} & \textcolor{gray}{4.77}
& 1.90 & \textcolor{red}{1.24} & \textcolor{gray}{1.48} & \textcolor{red}{1.24}
& \textcolor{red}{1.68} & \textcolor{green}{1.71} & \textcolor{gray}{1.72} & \textcolor{gray}{1.93}
& \textcolor{red}{2.55} & \textcolor{green}{2.65} & \textcolor{gray}{3.68} & \textcolor{gray}{3.60}
& 6.7 & \textcolor{red}{5.1} & \textcolor{gray}{7.7} & \textcolor{gray}{6.3} \\
& $V(B)$
& 4.06 & \textcolor{red}{1.94} & \textcolor{gray}{5.71} & \textcolor{gray}{2.74}
& 3.79 & \textcolor{red}{2.37} & \textcolor{gray}{3.62} & \textcolor{gray}{2.80}
& 0.88 & \textcolor{gray}{0.63} & \textcolor{gray}{1.34} & \textcolor{red}{0.44}
& \textcolor{red}{0.85} & \textcolor{gray}{0.91} & \textcolor{gray}{1.66} & \textcolor{green}{0.90}
& 2.39 & \textcolor{red}{1.64} & \textcolor{gray}{2.81} & \textcolor{gray}{1.83}
& 4.4 & \textcolor{red}{2.0} & \textcolor{gray}{5.7} & \textcolor{gray}{4.0} \\
& $V(R)$
& 2.40 & \textcolor{red}{0.95} & \textcolor{gray}{4.49} & \textcolor{gray}{1.17}
& 2.84 & \textcolor{red}{1.74} & \textcolor{gray}{2.73} & \textcolor{gray}{2.05}
& 1.29 & \textcolor{gray}{0.27} & \textcolor{gray}{0.85} & \textcolor{red}{0.18}
& \textcolor{red}{0.50} & \textcolor{green}{0.62} & \textcolor{gray}{1.01} & \textcolor{gray}{0.81}
& 1.21 & \textcolor{red}{1.12} & \textcolor{gray}{2.23} & \textcolor{gray}{1.86}
& 4.4 & \textcolor{red}{2.0} & \textcolor{gray}{5.8} & \textcolor{gray}{3.8} \\
\hline
512
& RMSE
& \textcolor{red}{3.59} & \textcolor{green}{3.98} & \textcolor{gray}{7.52} & \textcolor{gray}{4.08}
& \textcolor{red}{4.32} & \textcolor{gray}{5.15} & \textcolor{gray}{5.86} & \textcolor{green}{4.77}
& 1.19 & \textcolor{red}{0.85} & \textcolor{gray}{1.21} & \textcolor{gray}{0.90}
& \textcolor{red}{1.15} & \textcolor{gray}{1.64} & \textcolor{green}{1.46} & \textcolor{gray}{2.06}
& 3.03 & \textcolor{red}{2.76} & \textcolor{gray}{5.18} & \textcolor{gray}{8.76}
& 32.9 & \textcolor{red}{31.9} & \textcolor{gray}{35.2} & \textcolor{gray}{32.9} \\
& Bias
& \textcolor{red}{0.32} & \textcolor{gray}{3.02} & \textcolor{green}{0.94} & \textcolor{gray}{-1.74}
& 2.79 & \textcolor{gray}{4.27} & \textcolor{gray}{2.65} & \textcolor{red}{2.04}
& 0.73 & \textcolor{gray}{0.55} & \textcolor{red}{0.45} & \textcolor{gray}{0.47}
& 0.78 & \textcolor{gray}{1.33} & \textcolor{red}{0.49} & \textcolor{gray}{1.62}
& \textcolor{red}{0.27} & \textcolor{green}{0.78} & \textcolor{gray}{2.41} & \textcolor{gray}{7.99}
& -32.6 & \textcolor{red}{-31.7} & \textcolor{gray}{-34.8} & \textcolor{gray}{-32.6} \\
& Std
& 2.68 & \textcolor{red}{2.33} & \textcolor{gray}{4.40} & \textcolor{gray}{2.60}
& \textcolor{red}{2.13} & \textcolor{green}{2.62} & \textcolor{gray}{3.91} & \textcolor{gray}{3.01}
& 0.88 & \textcolor{red}{0.59} & \textcolor{gray}{0.85} & \textcolor{gray}{0.69}
& \textcolor{red}{0.76} & \textcolor{green}{0.91} & \textcolor{gray}{0.97} & \textcolor{gray}{0.96}
& 1.98 & \textcolor{red}{1.90} & \textcolor{gray}{3.00} & \textcolor{gray}{3.62}
& 4.1 & \textcolor{red}{3.7} & \textcolor{gray}{5.1} & \textcolor{gray}{4.4} \\
& $V(B)$
& 2.87 & \textcolor{red}{1.88} & \textcolor{gray}{7.28} & \textcolor{gray}{3.33}
& 4.43 & \textcolor{red}{2.26} & \textcolor{gray}{4.33} & \textcolor{gray}{3.65}
& 0.58 & \textcolor{red}{0.35} & \textcolor{gray}{0.92} & \textcolor{gray}{0.43}
& 0.69 & \textcolor{red}{0.62} & \textcolor{gray}{1.37} & \textcolor{gray}{1.11}
& 2.60 & \textcolor{red}{2.33} & \textcolor{gray}{5.43} & \textcolor{gray}{4.27}
& 4.4 & \textcolor{red}{1.8} & \textcolor{gray}{5.6} & \textcolor{gray}{6.1} \\
& $V(R)$
& \textcolor{red}{1.33} & \textcolor{green}{1.43} & \textcolor{gray}{3.84} & \textcolor{gray}{1.83}
& 3.53 & \textcolor{gray}{1.87} & \textcolor{gray}{2.52} & \textcolor{red}{1.75}
& 0.45 & \textcolor{red}{0.23} & \textcolor{gray}{0.52} & \textcolor{red}{0.23}
& 0.56 & \textcolor{red}{0.55} & \textcolor{gray}{0.90} & \textcolor{gray}{0.71}
& \textcolor{red}{1.05} & \textcolor{green}{1.28} & \textcolor{gray}{4.26} & \textcolor{gray}{4.60}
& 4.3 & \textcolor{red}{1.7} & \textcolor{gray}{5.5} & \textcolor{gray}{5.9} \\
\hline
1024
& RMSE
& 3.44 & \textcolor{red}{2.20} & \textcolor{gray}{8.26} & \textcolor{gray}{7.08}
& 4.41 & \textcolor{red}{2.60} & \textcolor{gray}{6.05} & \textcolor{gray}{4.40}
& 0.71 & \textcolor{red}{0.46} & \textcolor{gray}{0.66} & \textcolor{gray}{0.53}
& \textcolor{red}{0.96} & \textcolor{gray}{1.42} & \textcolor{green}{1.11} & \textcolor{gray}{2.04}
& 3.88 & \textcolor{red}{3.69} & \textcolor{gray}{6.82} & \textcolor{gray}{12.96}
& 34.0 & \textcolor{red}{32.5} & \textcolor{gray}{35.5} & \textcolor{gray}{33.6} \\
& Bias
& -2.94 & \textcolor{gray}{-1.54} & \textcolor{red}{-0.85} & \textcolor{gray}{-6.84}
& -2.47 & \textcolor{gray}{-0.82} & \textcolor{red}{0.69} & \textcolor{gray}{-2.19}
& 0.54 & \textcolor{gray}{0.36} & \textcolor{red}{0.22} & \textcolor{gray}{0.24}
& 0.87 & \textcolor{gray}{1.34} & \textcolor{red}{0.61} & \textcolor{gray}{1.41}
& \textcolor{red}{-0.71} & \textcolor{gray}{1.53} & \textcolor{green}{1.11} & \textcolor{gray}{12.30}
& -33.9 & \textcolor{red}{-32.4} & \textcolor{gray}{-35.2} & \textcolor{gray}{-33.5} \\
& Std
& 1.50 & \textcolor{red}{1.10} & \textcolor{gray}{3.91} & \textcolor{gray}{1.71}
& \textcolor{red}{1.35} & \textcolor{green}{1.53} & \textcolor{gray}{3.30} & \textcolor{gray}{2.43}
& 0.43 & \textcolor{red}{0.24} & \textcolor{gray}{0.46} & \textcolor{gray}{0.33}
& \textcolor{red}{0.36} & \textcolor{green}{0.45} & \textcolor{gray}{0.58} & \textcolor{gray}{0.67}
& \textcolor{red}{1.58} & \textcolor{green}{1.65} & \textcolor{gray}{3.16} & \textcolor{gray}{3.97}
& 2.7 & \textcolor{red}{2.4} & \textcolor{gray}{3.8} & \textcolor{gray}{3.3} \\
& $V(B)$
& 1.84 & \textcolor{red}{1.78} & \textcolor{gray}{8.50} & \textcolor{gray}{3.49}
& 4.20 & \textcolor{red}{2.45} & \textcolor{gray}{5.27} & \textcolor{gray}{3.86}
& 0.34 & \textcolor{red}{0.26} & \textcolor{gray}{0.48} & \textcolor{gray}{0.39}
& \textcolor{red}{0.45} & \textcolor{green}{0.47} & \textcolor{gray}{0.91} & \textcolor{gray}{1.57}
& 4.05 & \textcolor{red}{3.97} & \textcolor{gray}{8.50} & \textcolor{gray}{7.51}
& 3.6 & \textcolor{red}{1.6} & \textcolor{gray}{6.1} & \textcolor{gray}{7.2} \\
& $V(R)$
& 1.50 & \textcolor{red}{1.32} & \textcolor{gray}{4.46} & \textcolor{gray}{3.44}
& 2.20 & \textcolor{red}{1.36} & \textcolor{gray}{2.27} & \textcolor{gray}{2.49}
& 0.30 & \textcolor{gray}{0.21} & \textcolor{gray}{0.19} & \textcolor{red}{0.18}
& \textcolor{red}{0.40} & \textcolor{green}{0.44} & \textcolor{gray}{0.51} & \textcolor{gray}{0.80}
& \textcolor{red}{1.76} & \textcolor{green}{2.50} & \textcolor{gray}{5.92} & \textcolor{gray}{7.85}
& 3.6 & \textcolor{red}{1.6} & \textcolor{gray}{6.1} & \textcolor{gray}{7.1} \\
\hline
2048
& RMSE
& \textcolor{red}{9.55} & \textcolor{gray}{10.44} & \textcolor{green}{9.76} & \textcolor{gray}{15.13}
& 8.64 & \textcolor{gray}{6.71} & \textcolor{red}{6.34} & \textcolor{gray}{7.88}
& 0.44 & \textcolor{red}{0.34} & \textcolor{gray}{0.45} & \textcolor{gray}{0.39}
& 1.16 & \textcolor{gray}{1.54} & \textcolor{red}{1.02} & \textcolor{gray}{2.11}
& \textcolor{red}{4.95} & \textcolor{green}{5.07} & \textcolor{gray}{9.22} & \textcolor{gray}{18.87}
& 34.7 & \textcolor{red}{33.4} & \textcolor{gray}{35.2} & \textcolor{gray}{34.1} \\
& Bias
& -9.50 & \textcolor{gray}{-10.42} & \textcolor{red}{-6.85} & \textcolor{gray}{-15.08}
& -8.41 & \textcolor{gray}{-6.63} & \textcolor{red}{-3.67} & \textcolor{gray}{-7.42}
& 0.35 & \textcolor{gray}{0.31} & \textcolor{red}{0.03} & \textcolor{gray}{0.18}
& 1.14 & \textcolor{gray}{1.51} & \textcolor{red}{0.73} & \textcolor{gray}{1.46}
& -1.33 & \textcolor{gray}{3.25} & \textcolor{red}{0.31} & \textcolor{gray}{18.39}
& -34.7 & \textcolor{red}{-33.4} & \textcolor{gray}{-35.0} & \textcolor{gray}{-34.0} \\
& Std
& 0.97 & \textcolor{red}{0.70} & \textcolor{gray}{2.64} & \textcolor{gray}{1.28}
& \textcolor{red}{0.91} & \textcolor{green}{1.01} & \textcolor{gray}{2.43} & \textcolor{gray}{1.74}
& 0.22 & \textcolor{red}{0.13} & \textcolor{gray}{0.26} & \textcolor{gray}{0.19}
& \textcolor{red}{0.22} & \textcolor{green}{0.25} & \textcolor{gray}{0.36} & \textcolor{gray}{0.53}
& \textcolor{red}{1.24} & \textcolor{green}{1.48} & \textcolor{gray}{3.20} & \textcolor{gray}{3.99}
& \textcolor{red}{1.7} & \textcolor{red}{1.7} & \textcolor{gray}{3.1} & \textcolor{gray}{2.7} \\
& $V(B)$
& 2.19 & \textcolor{red}{1.54} & \textcolor{gray}{8.68} & \textcolor{gray}{3.60}
& 3.83 & \textcolor{red}{2.24} & \textcolor{gray}{5.75} & \textcolor{gray}{3.58}
& 0.27 & \textcolor{red}{0.18} & \textcolor{gray}{0.42} & \textcolor{gray}{0.36}
& \textcolor{red}{0.32} & \textcolor{green}{0.38} & \textcolor{gray}{0.79} & \textcolor{gray}{1.65}
& 5.59 & \textcolor{red}{5.54} & \textcolor{gray}{11.34} & \textcolor{gray}{11.50}
& 3.6 & \textcolor{red}{1.3} & \textcolor{gray}{6.8} & \textcolor{gray}{7.6} \\
& $V(R)$
& 2.17 & \textcolor{red}{1.53} & \textcolor{gray}{5.73} & \textcolor{gray}{3.61}
& 3.34 & \textcolor{red}{2.22} & \textcolor{gray}{3.60} & \textcolor{gray}{2.90}
& 0.21 & \textcolor{red}{0.16} & \textcolor{gray}{0.18} & \textcolor{gray}{0.21}
& \textcolor{red}{0.31} & \textcolor{green}{0.37} & \textcolor{gray}{0.45} & \textcolor{gray}{0.83}
& \textcolor{red}{2.76} & \textcolor{green}{4.04} & \textcolor{gray}{7.03} & \textcolor{gray}{11.63}
& 3.6 & \textcolor{red}{1.2} & \textcolor{gray}{6.7} & \textcolor{gray}{7.5} \\
\hline
\end{tabular}
}
\end{table*}

\begin{figure}[h]
\centering
\includegraphics[angle = 0, width=0.9\linewidth]{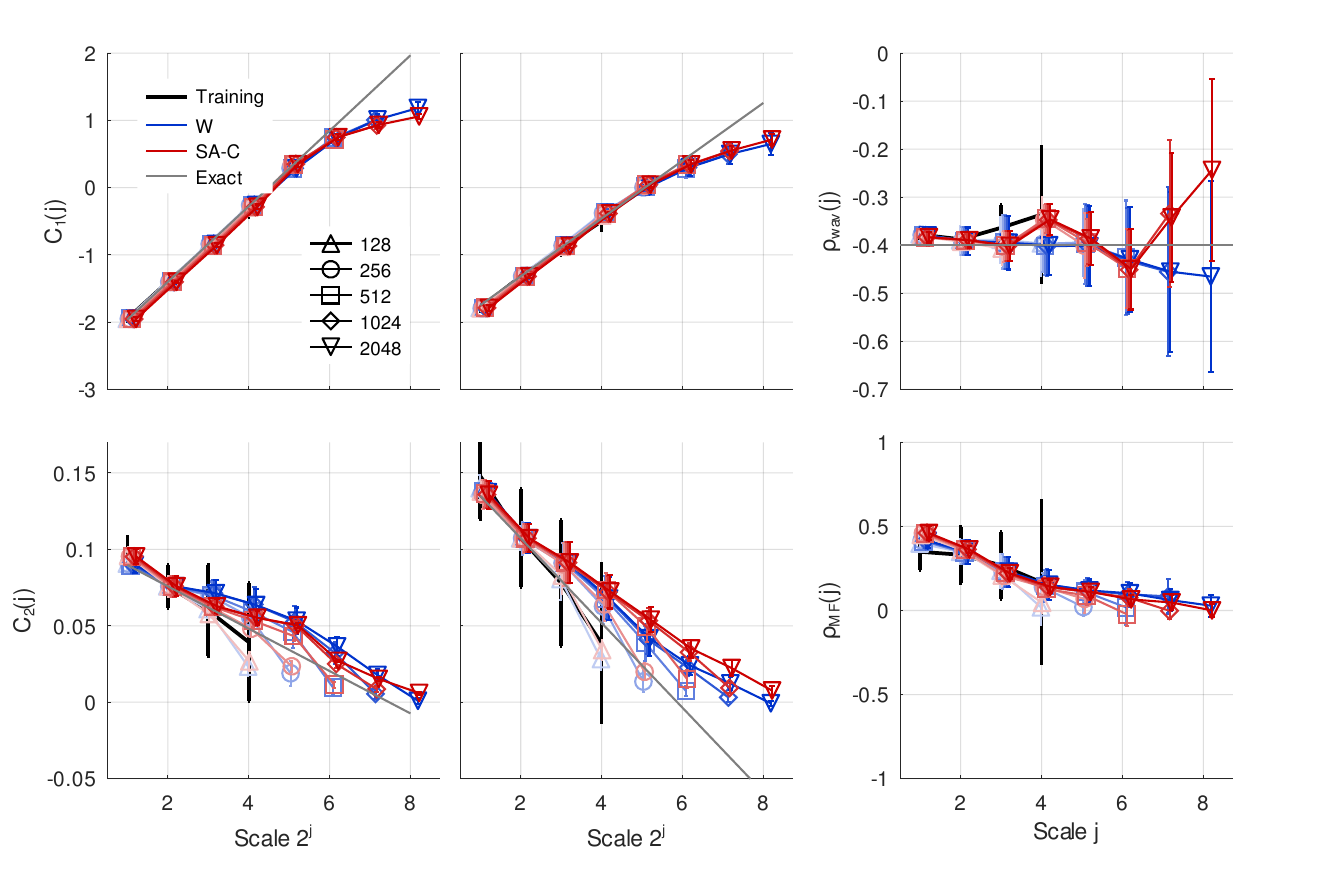}
\vskip-5mm
\caption{ Scale-wise comparison of W and SA-C for the different generated texture sizes.}
\label{fig:multiscale}
\end{figure}

\noindent\textbf{Architecture design and complexity.}
\KAKA{A comparison} of $3\times3$, $9\times9$, and $15\times15$ first-layer convolutional kernel sizes showed that the $3\times3$ kernel provided the best overall validation performance while requiring the lowest parameter and computational costs.
\KAKA{Detailed results are omitted due to space constraints, and all reported performances use the $3\times3$ kernel.}
\KAKA{This indicates that a small receptive field is sufficient to reproduce local bursty bivariate structures}.\\
\noindent\textbf{Spatial-support extrapolation performance.}
Table~\ref{tab:performance} compares the statistical performance of the four attention configurations over the different generated texture sizes.
Among the $30$ combinations of output size and statistical descriptor, SA-C achieves the lowest RMSE in $15$ cases and ties for the lowest RMSE in one additional case.
Its advantage is particularly clear at $N=1024$, where it provides the lowest RMSE for five of the six descriptors.
The models generate textures up to $N=1024$, eight times the training size along each spatial dimension, with good overall agreement across the evaluated descriptors. At $N=2048$, the RMSE increases for the first-order scaling
coefficients and the wavelet correlation in all configurations.
Attention in the generator does not provide the same improvement.
\KAKA{SA-G presents larger variability across independent trainings and degrades the scaling and cross-correlation statistics as the generated size increases, particularly from $N=1024$.}
\KAKA{A possible explanation is that generator attention is trained on $16\times16$ feature maps but is applied to increasingly larger feature maps during extrapolation. The resulting size-dependent mixing may distort the scaling and cross-component correlations. In contrast, critic attention is used only during training at the fixed training size and does not modify the generated features at inference.}\\
Beyond the aggregate errors, Fig.~\ref{fig:multiscale} compares the scale-by-scale statistical behavior of W and SA-C.
Both configurations reproduce the approximately linear evolution of $C_1(j)$ and $C_2(j)$ \KAKA{up to $N=1024$. The deviations observed at $N=2048$ confirm the extrapolation limit identified in Table~\ref{tab:performance}.}\\

\begin{figure}[!t]
\centering

\begin{subfigure}[t]{0.49\linewidth}
    \centering
    \makebox[\linewidth][c]{%
        \includegraphics[width=0.06\linewidth]
        {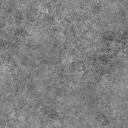}%
        \hspace{0.005\linewidth}%
        \includegraphics[width=0.06\linewidth]
        {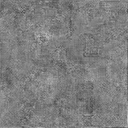}%
        \hspace{0.005\linewidth}%
        \includegraphics[width=0.12\linewidth]
        {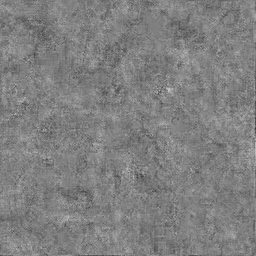}%
        \hspace{0.005\linewidth}%
        \includegraphics[width=0.24\linewidth]
        {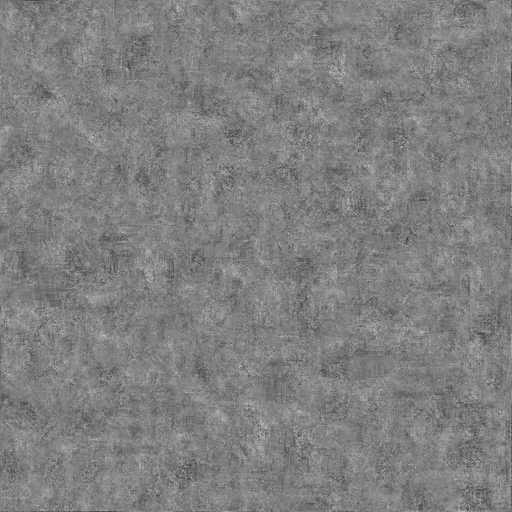}%
        \hspace{0.005\linewidth}%
        \includegraphics[width=0.48\linewidth]{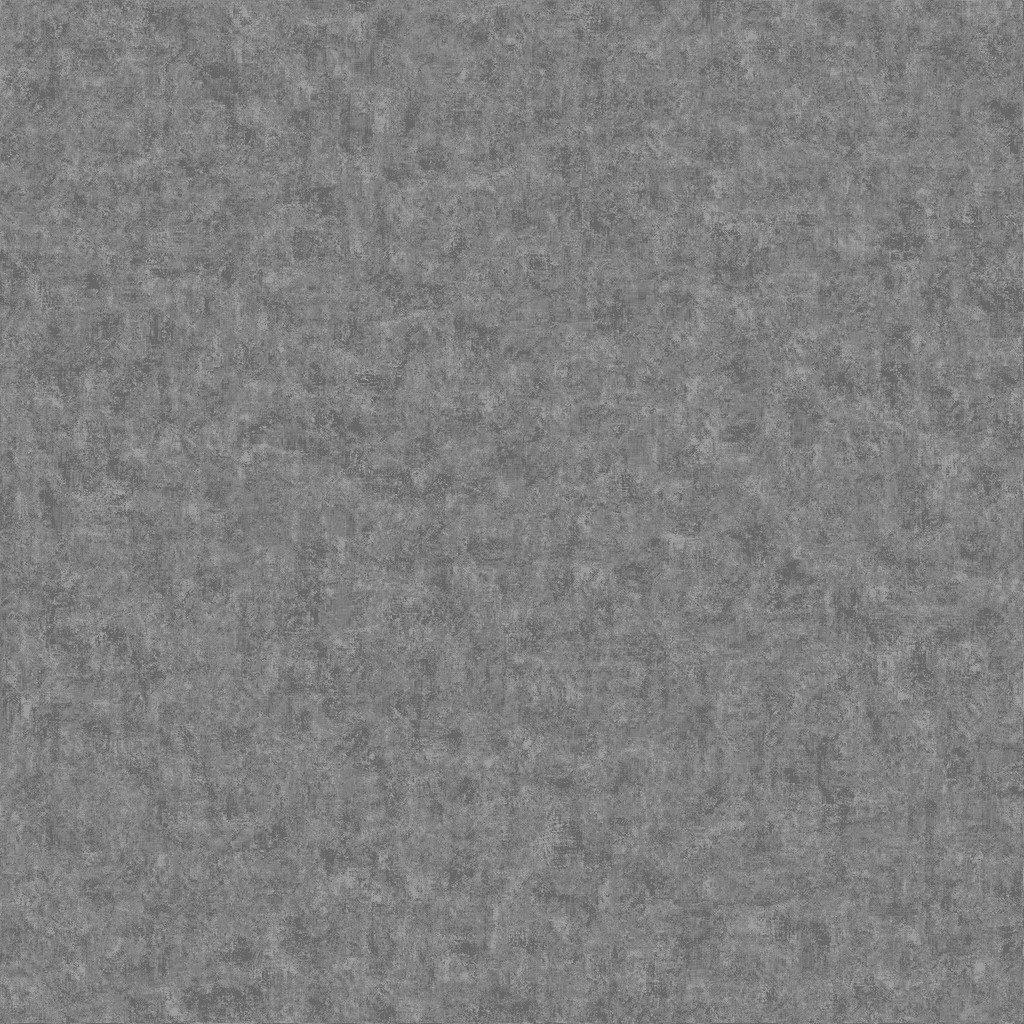}%
    }
    \caption{$H=0.8$.}
    \label{fig:generated_H06}
\end{subfigure}
\hfill
\begin{subfigure}[t]{0.49\linewidth}
    \centering
    \makebox[\linewidth][c]{%
        \includegraphics[width=0.06\linewidth]
        {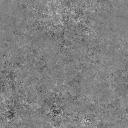}%
        \hspace{0.005\linewidth}%
        \includegraphics[width=0.06\linewidth]
        {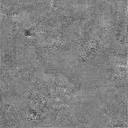}%
        \hspace{0.005\linewidth}%
        \includegraphics[width=0.12\linewidth]
        {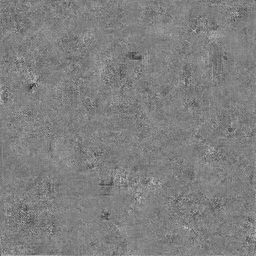}%
        \hspace{0.005\linewidth}%
        \includegraphics[width=0.24\linewidth]
        {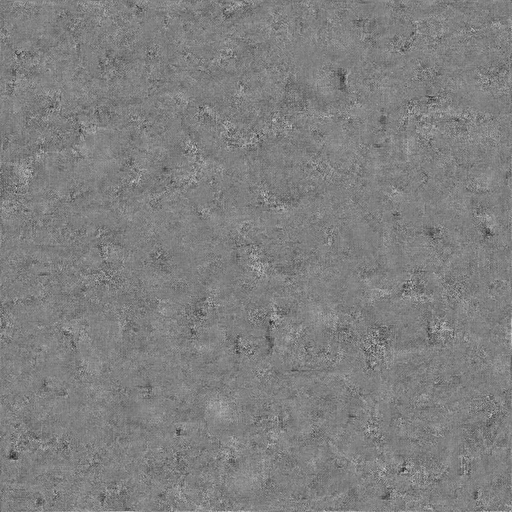}%
        \hspace{0.005\linewidth}%
        \includegraphics[width=0.48\linewidth]
        {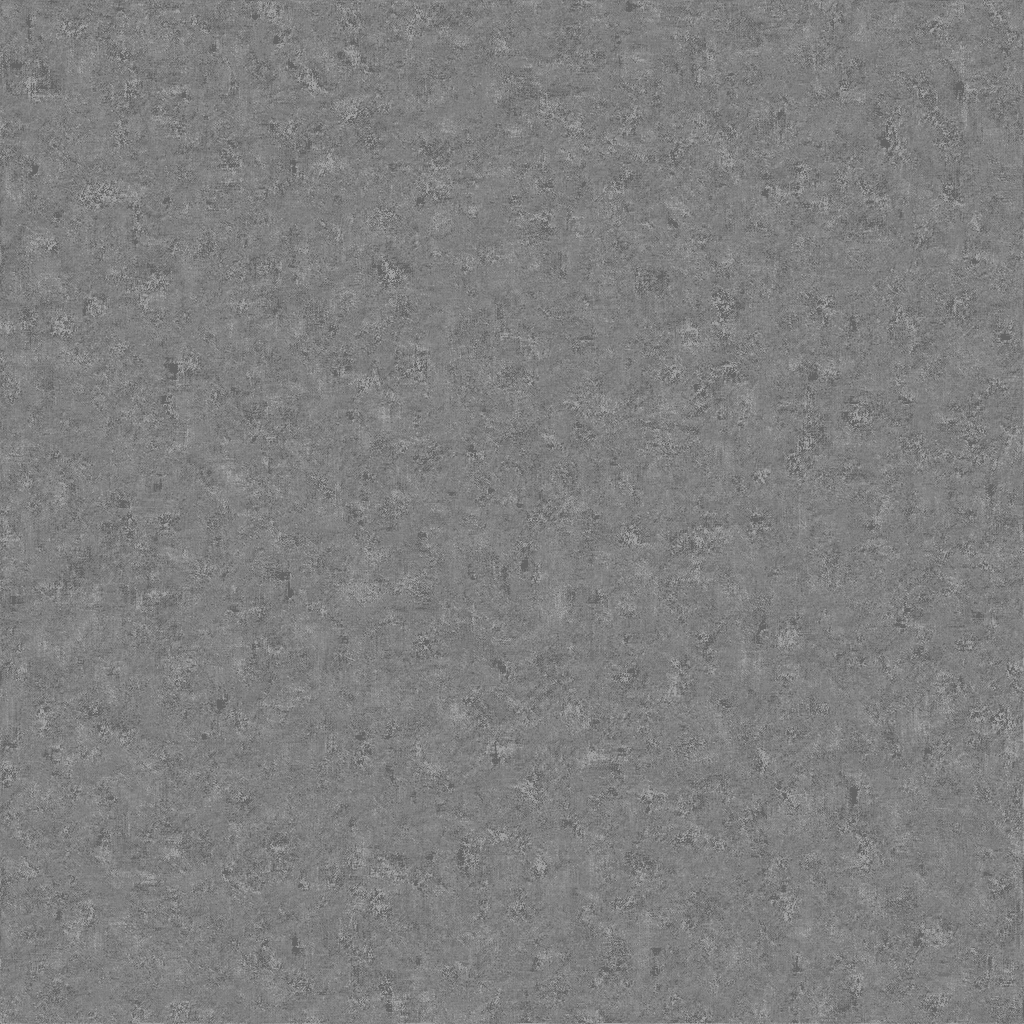}%
    }
    \caption{$H=0.6$.}
    \label{fig:generated_H08}
\end{subfigure}

\vspace{-2mm}
\caption{Training textures at $N_T=128$ followed by SA-C-generated
textures at $N=128$, $256$, $512$, and $1024$, from left to right,
for the two components with $H=0.8$ and $H=0.6$.}
\label{fig:generated_examples}
\end{figure}
Finally, Fig.~\ref{fig:generated_examples} shows that the SA-C generator trained at $N_T=128$ produces visually consistent textures at increasing spatial supports.



\section{Conclusions and perspectives}
\label{sec:Conclusion}
This work addressed the generation of multivariate multifractal scale-free textures on spatial supports larger than those available during training.
\KAKA{A fully convolutional WGAN trained on $128\times128$ textures reproduces the targeted multiscale and cross-statistical properties up to $1024\times1024$, without including these properties in the training loss.}
\KAKA{Among the evaluated self-attention placements, critic-only attention provides the most consistent performance while introducing no additional cost during generation. Performance degradation at $2048\times2048$ indicates the current extrapolation limit.}
\KAKA{Future work will consider higher-dimensional textures and statistically informed learning objectives.}

\clearpage
{\ninept
\bibliographystyle{IEEEbib}
\bibliography{strings,refs}
}




\end{document}